\documentclass[12pt,prd,preprint,tightenlines,floatfix,
showpacs,preprintnumbers,nofootinbib,eqsecnum,superscriptaddress,final]%
{revtex4}

\usepackage{amsmath,amsfonts,amssymb,amstext,mathrsfs}
\usepackage[dvips]{graphicx}

\begin{document}

\title{Double parton effects for jets with large rapidity separation}


\author{Antoni Szczurek}
 \email{Antoni.Szczurek@ifj.edu.pl}
\affiliation{Institute of Nuclear Physics PAN, PL-31-342 Cracow, Poland}
\affiliation{University of Rzesz\'ow, PL-35-959 Rzesz\'ow, Poland}

\author{Anna Cisek}
  \affiliation{University of Rzesz\'ow, Poland}

\author{Rafal Maciu{\l}a}
  \affiliation{Institute of Nuclear Physics PAN, Krak\'ow, Poland}

\begin{abstract}
We discuss production of four jets $pp \rightarrow j j j j X$
with at least two jets with large rapidity separation in proton-proton
collisions at the LHC through the mechanism of double-parton scattering 
(DPS).
The cross section is calculated in a factorizaed approximation.
Each hard subprocess is calculated in LO collinear approximation.  
The LO pQCD calculations are shown to give a reasonably good
descritption of CMS and ATLAS data on inclusive jet production.
It is shown that relative contribution of DPS is growing with increasing
rapidity distance between the most remote jets, center-of-mass energy
and with decreasing (mini)jet transverse momenta.
We show also result for angular azimuthal dijet correlations calculated 
in the framework of $k_t$-factorization approaximation.
\end{abstract}


\maketitle


\section{Introduction}

\begin{figure}[!h]
\includegraphics[width=6cm]{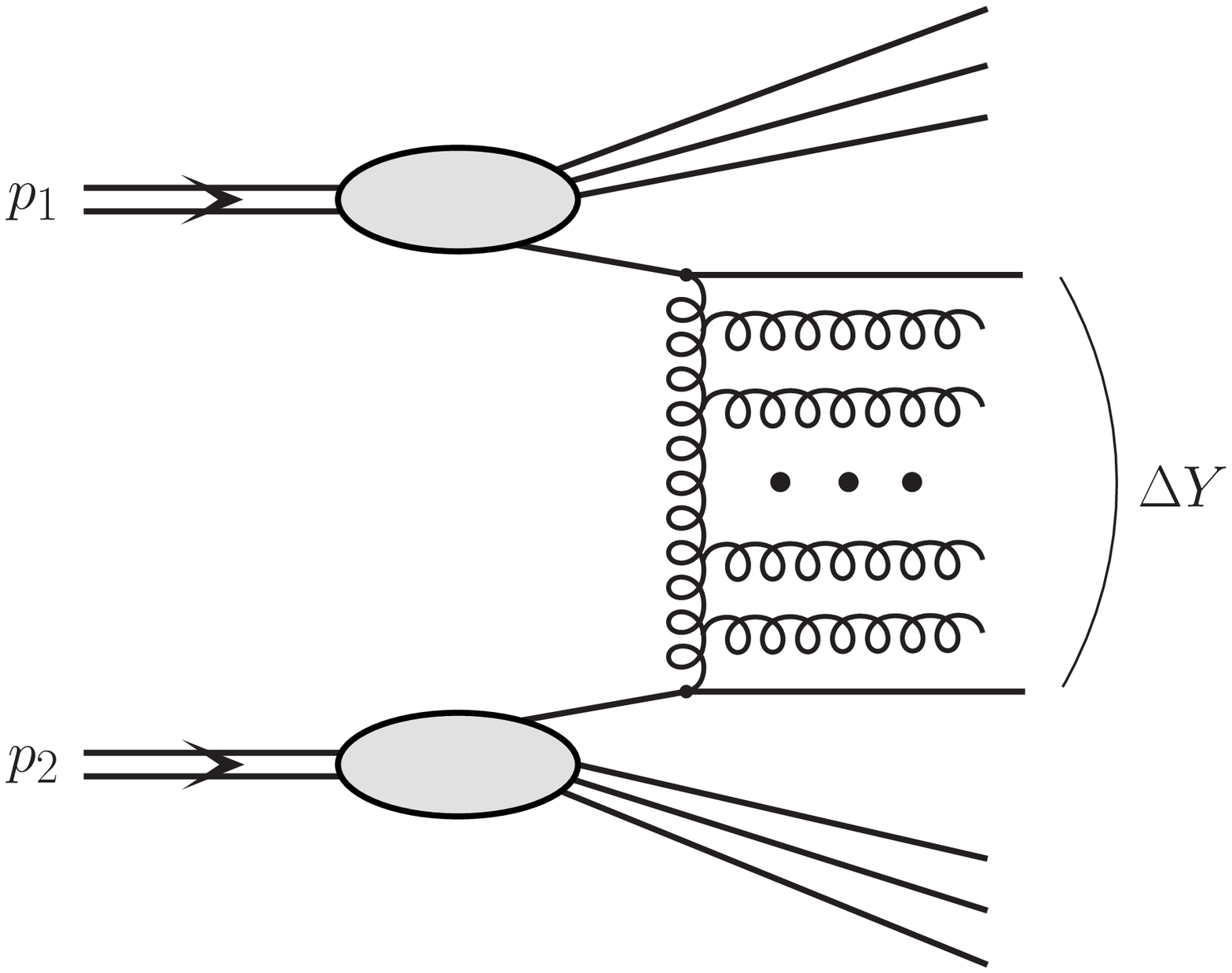}
\includegraphics[width=6cm]{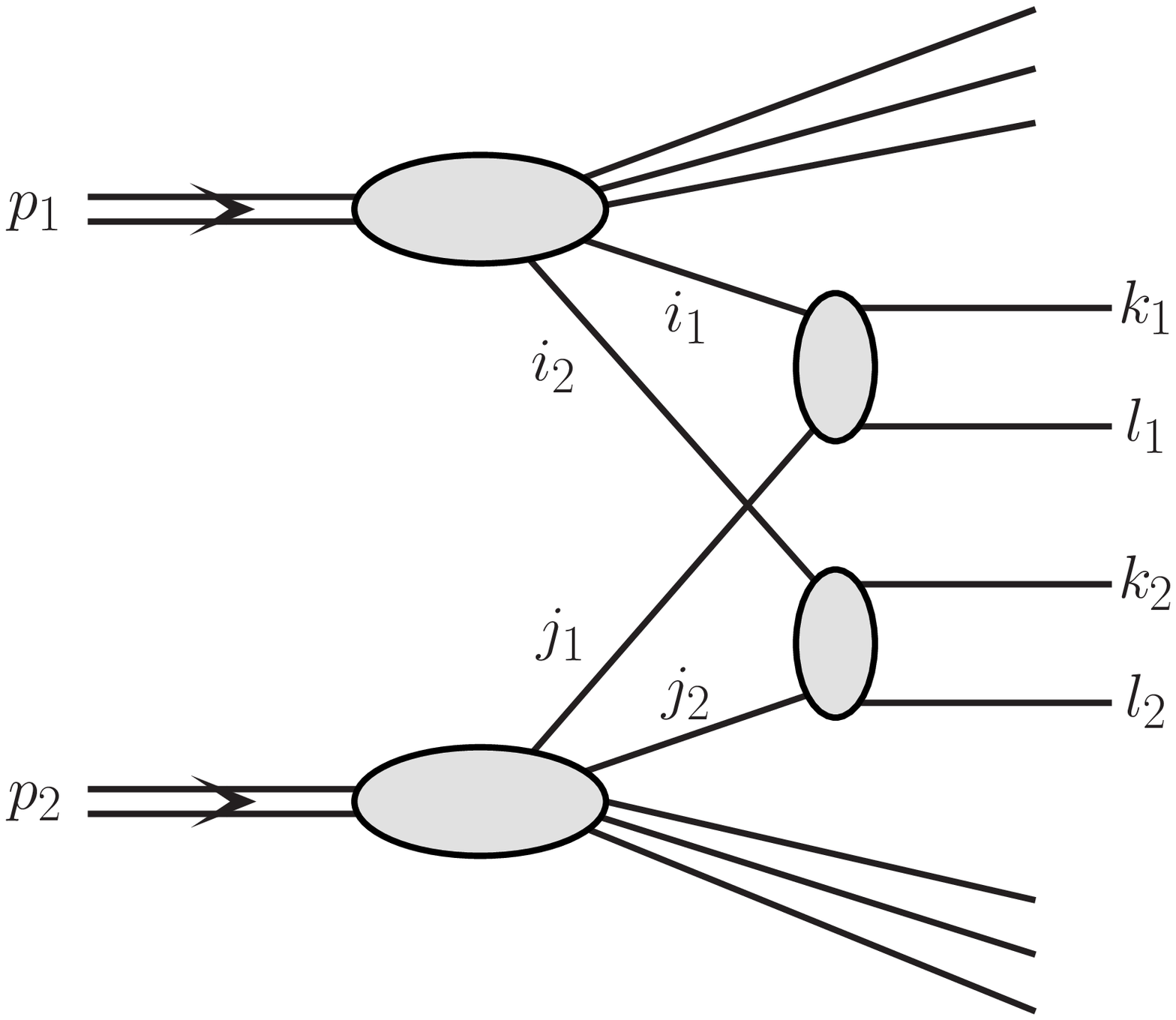}
   \caption{
\small The production of jets widely separated in rapidity via
Mueller-Navelet mechanism (left) and within double-parton scattering 
mechanism (right).
}
 \label{fig:diagrams}
\end{figure}

Almost 25 years ago Mueller and Navelet predicted strong decorrelation 
in relative azimuthal angle \cite{Mueller:1986ey} of jets widely
separated in rapidity due to exchange of the BFKL ladder between
quarks. The underlying mechanism is shown
in the left panel of Fig.\ref{fig:diagrams}.
The leading-logarithm approximation was studied in several papers
\cite{Mueller:1986ey,DelDuca:1993mn,Stirling:1994he,DelDuca:1994ng,Kim96,
Andersen2001}. 
Recent higher-order BFKL calculation slightly modified
this simple picture \cite{Bartels-MNjets,Vera:2007kn,Marquet:2007xx,
Colferai:2010wu,Caporale:2011cc,Ivanov:2012ms,
Caporale:2012ih,Ducloue:2013hia,Ducloue:2013bva,DelDuca2014}.
The involved calculations give similar result as the standard
next-to-leading order perturbative calculations in QCD.

On the other hand recent studies of multiparton interactions shown
that such effects may easily produce two objects (particles) that 
are emitted far in rapidity.
Very good example is production of $c \bar c c \bar c$ 
\cite{Luszczak:2011zp,Maciula:2013kd,Hameren2014}
or inclusive production of two $J/\psi$ 
\cite{Kulesza-Stirling,Baranov:2012re}.
Recently two of us presented a first estimation of the DPS effects
for jets widely separated in rapidity \cite{MS2014}. The underlying mechanism
is shown in the right panel of Fig.\ref{fig:diagrams}. 

\begin{figure}
\includegraphics[width=5cm]{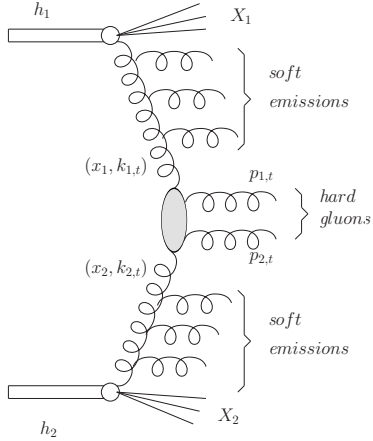}
  \caption{
\small A typical diagram with two gluonic ladders relevant for 
the $k_t$-factorization approach.
}
 \label{fig:diagram_kt_factorization}
\end{figure}

Here we report first results for DPS production of jets widely separated
in rapidity. In addition we shall show first corresponding
results obtained within $k_t$-factorization approach \cite{CMS2014}.
An example of corresponding diagram in show 
in Fig.\ref{fig:diagram_kt_factorization}. The blob in the figure represents
hard off-shell matrix element. As an example we shall discuss azimuthal 
correlation between jets from the blob. This is very different than
in the BFKL Mueller-Navelet approach where correlation between the most
remote (mini)jet in the ladder (see left panel) are considered.

In this presentation we shall compare the different results and discuss
some possibile future directions.

\section{Basics of the formalism}

All partonic cross sections are calculated only in leading order.
The cross section for dijet production can be then written as:
\begin{equation}
\frac{d \sigma(i j \to k l)}{d y_1 d y_2 d^2p_t} 
= \frac{1}{16 \pi^2 {\hat s}^2}
\sum_{i,j} x_1 f_i(x_1,\mu^2) \; x_2 f_j(x_2,\mu^2) \;
\overline{|\mathcal{M}_{i j \to k l}|^2} \;.
\label{LO_SPS}
\end{equation}
In our calculations we include all leading-order $i j \to k l$ partonic 
subprocesses.
The $K$-factor for dijet production are rather small, of the order of 
$1.1 - 1.3$ (see e.g. \cite{K-factor1,K-factor2}, 
but can be easily incorporated in our calculations. Below we shall show that
already the leading-order approach gives results in sufficiently reasonable 
agreement with recent ATLAS \cite{ATLASjets} and CMS \cite{CMSjets} data.

This simplified leading-order approach can be however used easily in 
calculating DPS differential cross sections. 
A multi-dimensional cross section can be written as:
\begin{equation}
\frac{d \sigma^{DPS}(p p \to \textrm{4jets} \; X)}{d y_1 d y_2 d^2p_{1t}
  d y_3 d y_4 d^2p_{2t}} 
= \sum_{i_1,j_1,k_1,l_1;i_2,j_2,k_2,l_2} \; 
\frac{\mathcal{C}}{\sigma_{eff}} \;
\frac{d \sigma(i_1 j_1 \to k_1 l_1)}{d y_1 d y_2 d^2p_{1t}} \; 
\frac{d \sigma(i_2 j_2 \to k_2 l_2)}{d y_3 d y_4 d^2p_{2t}}\;, 
\label{DPS}
\end{equation}
where
$\mathcal{C} = \left\{ \begin{array}{ll}
\frac{1}{2}\;\; & \textrm{if} \;\;i_1 j_1 = i_2 j_2 \wedge k_1 l_1 = k_2 l_2\\
1\;\;           & \textrm{if} \;\;i_1 j_1 \neq i_2 j_2 \vee k_1 l_1 \neq k_2 l_2
\end{array} \right\} $ and partons $j,k,l,m = g, u, d, s, \bar u, \bar d, \bar s$. 
The combinatorial factors include identity of the two subprocesses.
Each step of DPS is calculated in the leading-order approach 
(see Eq.(\ref{LO_SPS})).

Experimental data from Tevatron \cite{Tevatron} and 
LHC \cite{Aad:2013bjm} 
provide an estimate of $\sigma_{eff}$ in the denominator of formula 
(\ref{DPS}). In the calculations, result of which are presented here, 
we have taken $\sigma_{eff}$ = 15 mb.

In this presentation we shall also show some correlations between
jets obtained in the $k_t$-factorization approach.
In the $k_t$-factorization approach the cross section differential in 
each jet rapidity and transverse momentum can be written as:
\begin{eqnarray}
\frac{d \sigma}{d y_1 d y_{2} d^2 p_{1,t} d^2 p_{2,t}} =
\sum_{i,j,k,l} \; \int \frac{d^2 \kappa_{1,t}}{\pi} 
\frac{d^2 \kappa_{2,t}}{\pi}
\frac{1}{16 \pi^2 (x_1 x_2 s)^2} \; 
\overline{ | {\cal M}_{ij \rightarrow k l} |^2} 
\;\;\;\;\;\;\;\;\;\;\;\;\;\; \nonumber \\ 
\;\;\;\;\;\;\;\;\;\;\;\;\;\;\;\;\;\;\;\;\;\;\;\;\;\;\;\;\;\;\;\;\;\;\;\;\; \times \;\;\;
\delta^{2} \left( \vec{\kappa}_{1,t} + \vec{\kappa}_{2,t} - \vec{p}_{1,t} - \vec{p}_{2,t} \right) \;
{\cal F}_i(x_1,\kappa_{1,t}^2,\mu_f^2) \; 
{\cal F}_j(x_2,\kappa_{2,t}^2,\mu_f^2) \; .   
\label{kt_factorization_formula}
\end{eqnarray}
Above ${\cal F}$'s denote unintegrated gluon (parton) distribution (UGDF).
The indecies $i,j$ mean reggeized partons, gluons, quarks and
antiquarks. 
In this presentation we shall use Kimber-Martin-Ryskin unintegrated
parton distributions \cite{KMR}. 
The formulae for off-shell matrix elements were obtained e.g. 
in \cite{NSS2013} and corresponding formulae can be used in our calculations.

\section{Selected results}

\begin{figure}[!h]
\includegraphics[width=5.5cm]{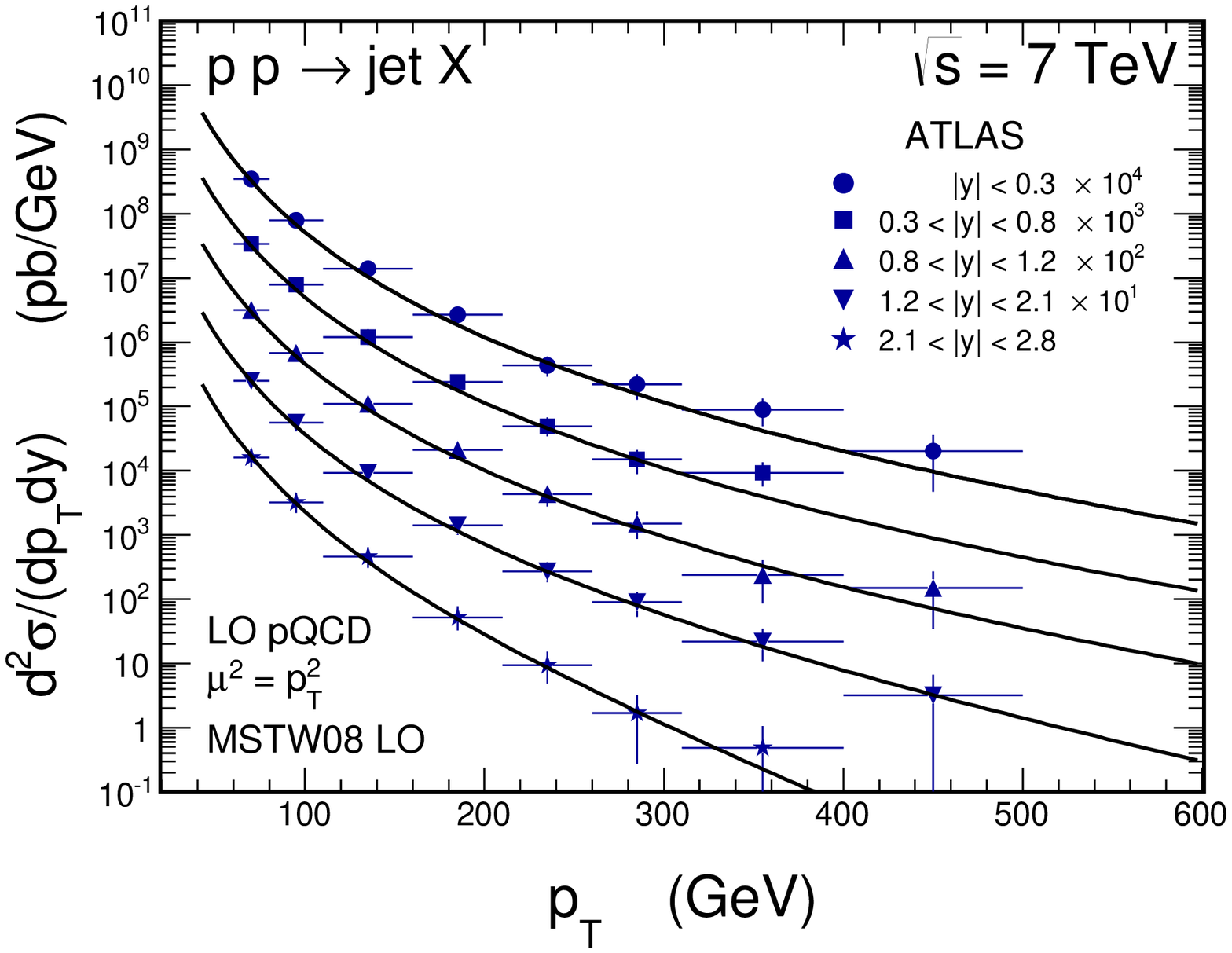}
\includegraphics[width=5.5cm]{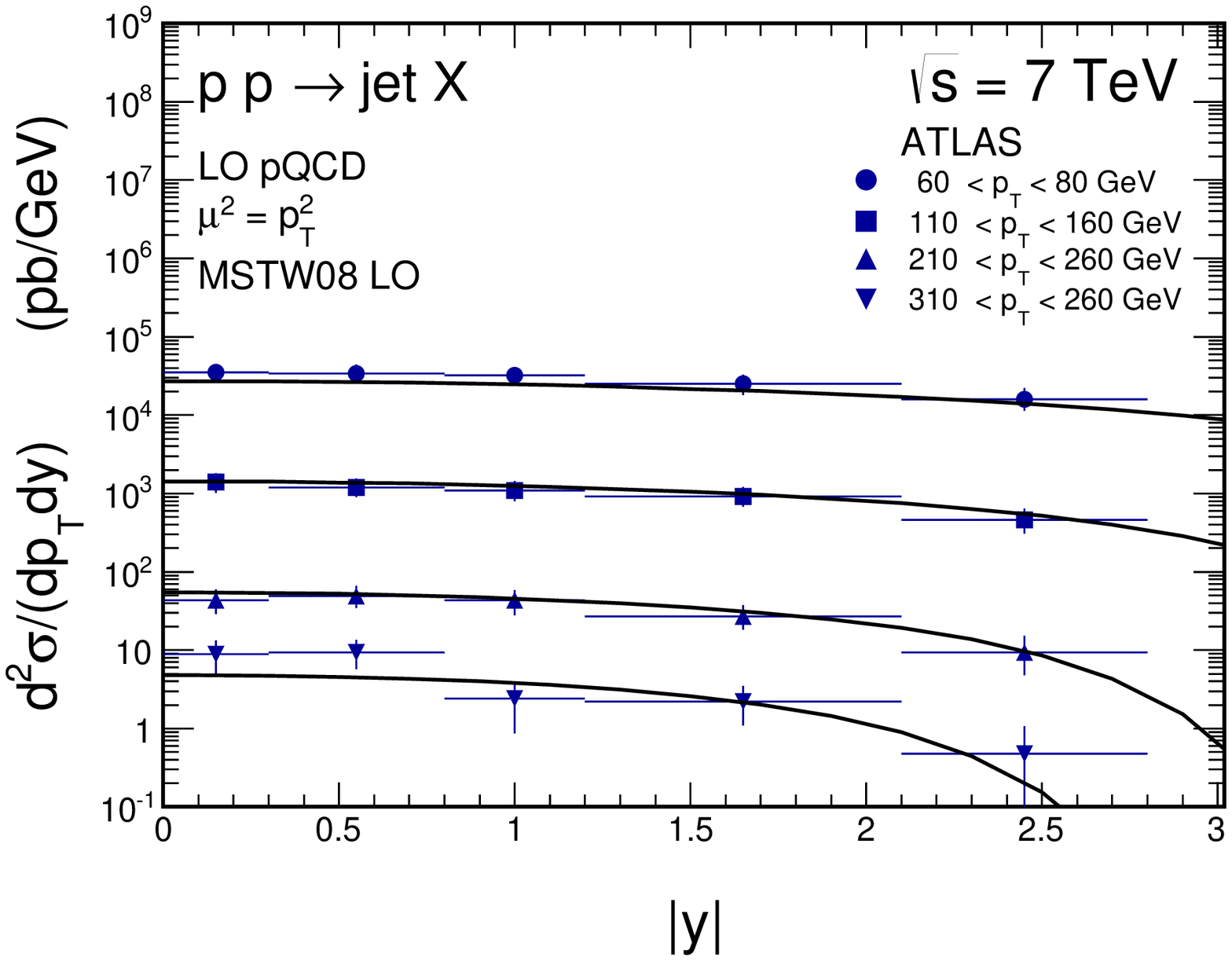}
   \caption{
\small Transverse momentum distribution of jets for different regions
of the jet rapidity (left panel) and corresponding rapidity distribution
of jets with different cuts in $p_t$ as specified in the figure
caption. The theoretical calculations were performed with 
the MSTW08 parton distributions \cite{MSTW08}.
The data points were obtained by the ATLAS collaboration \cite{ATLASjets}.
}
 \label{fig:pt-and-y-spectra-ATLASjets}
\end{figure}

Before we shall show our results for rapidity-distant-jet correlations 
we wish to show the quality of the description of some observables 
for inclusive jet production. In Fig.~\ref{fig:pt-and-y-spectra-ATLASjets}
we show as an example distributions in jet transverse momentum for different
intervals of jet (pseudo)rapidity. In these calculations we have used
MSTW08 PDFs \cite{MSTW08}. The agreement with recent ATLAS data
\cite{ATLASjets} is fairly reasonable which allows to use the same 
distributions for the evalution of the DPS effects for large rapidity 
distances between jets.
In Ref.\cite{MS2014} we have shown also some results relevant for the CMS
collaboration measurements. 

\begin{figure}[!h]
\includegraphics[width=5cm]{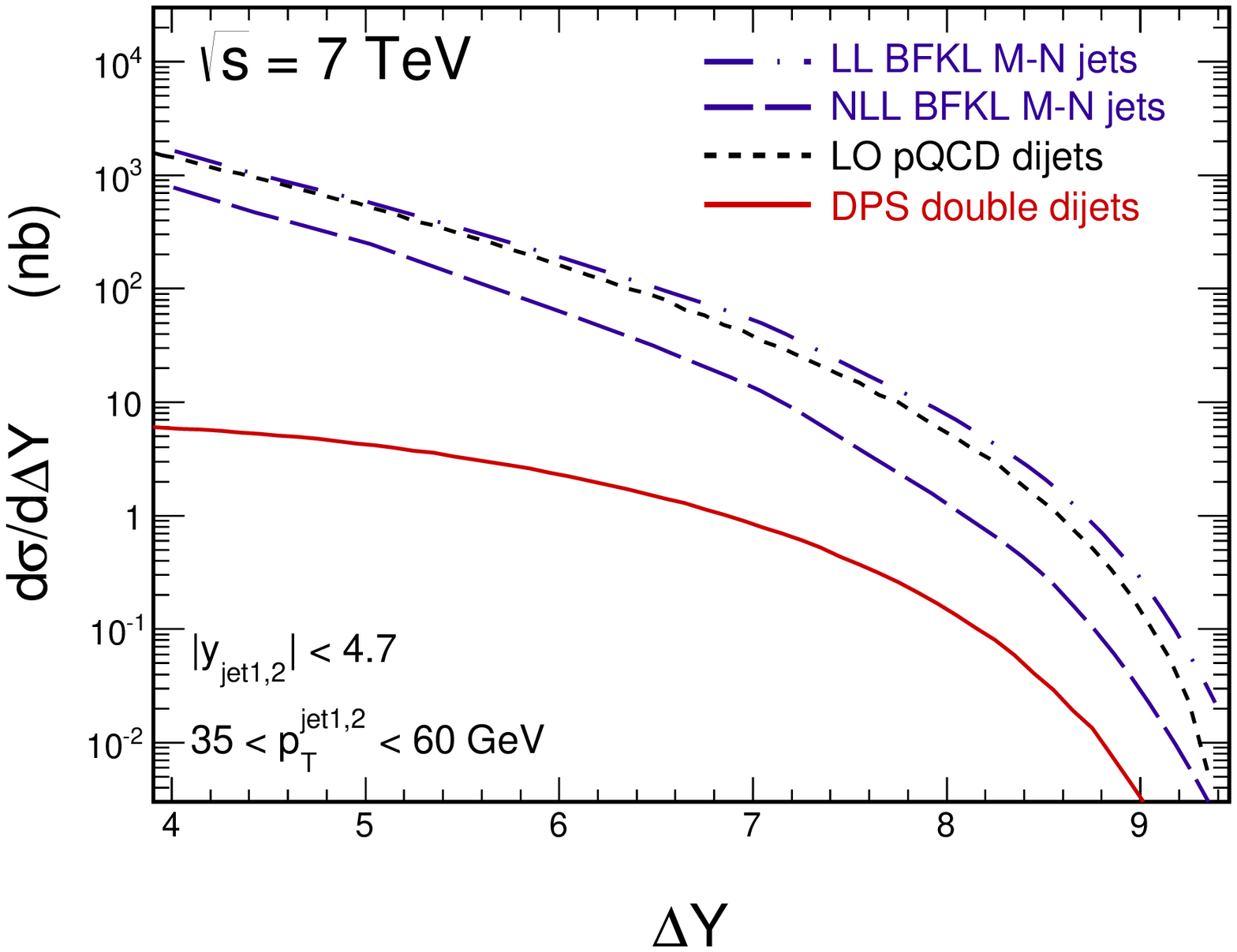}
\includegraphics[width=5cm]{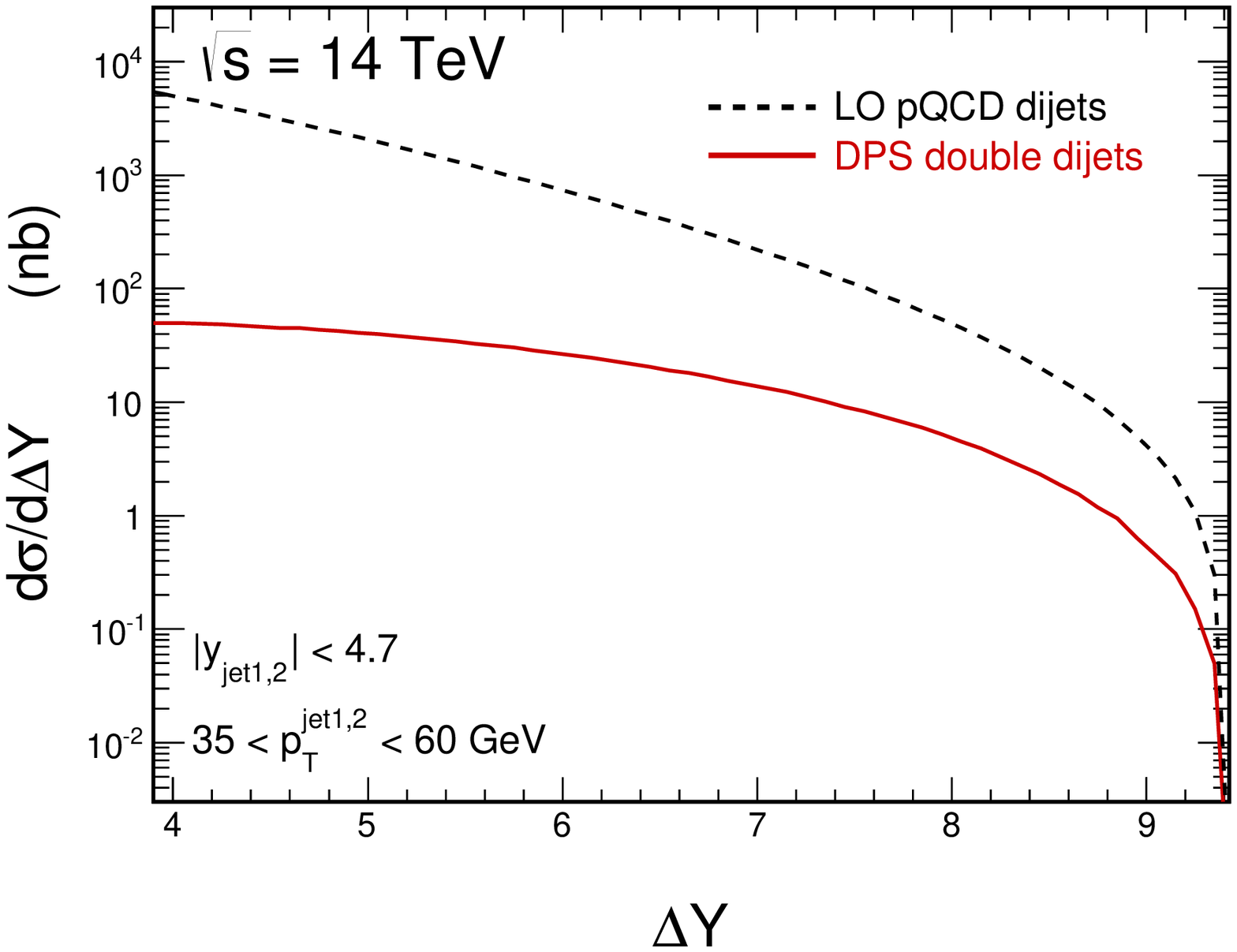}
   \caption{
\small Distribution in rapidity distance between the jet 
(35 GeV $< p_t <$ 60 GeV) with maximal (the most positive) and minimal 
(the most negative) rapidities. 
The collinear pQCD result is shown by the short-dashed line
and the DPS result by the solid line for $\sqrt{s}$ = 7 TeV (left panel)
and $\sqrt{s}$ = 14 TeV (right panel). For comparison we also show
results for the BFKL Mueller-Navelet jets in leading-logarithm 
and next-to-leading-order logarithm approaches from 
Ref.~\cite{Ducloue:2013hia}.
}
 \label{fig:Deltay1}
\end{figure}

Now we shall proceed to the jets with large rapidity separation.
In Fig.~\ref{fig:Deltay1} we show distribution in the rapidity 
distance between two jets in leading-order collinear calculation
and between the most distant jets in rapidity in the case of four DPS jets.
In this calculation we have included cuts characteristic for the
CMS expriments \cite{CMS_private}.
For comparison we show also results for BFKL calculation from
Ref.~\cite{Ducloue:2013hia}. For this kinematics the DPS jets
give sizeable contribution only at large rapidity distance.
However, the four-jet (DPS) and dijet (LO SPS) final state can 
be easily distinguished and in principle one can concentrate on the
DPS contribution which is interesting by itself.
The NLL BFKL cross section (long-dashed line) is sizeably lower than 
that for the LO collinear approach (short-dashed line).
This causes that the general situation is not clear.

\begin{figure}[!h]
\includegraphics[width=4cm]{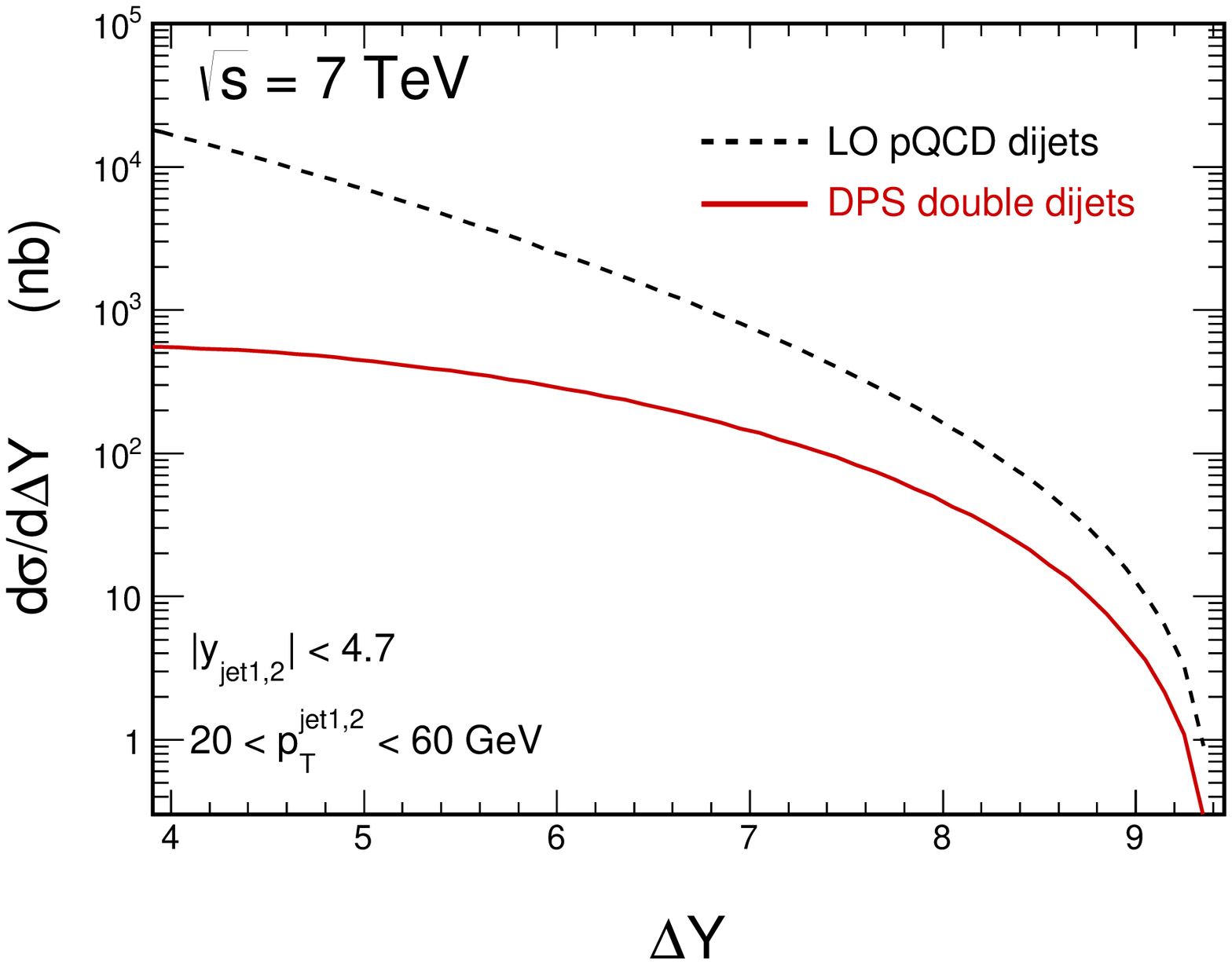}
\includegraphics[width=4cm]{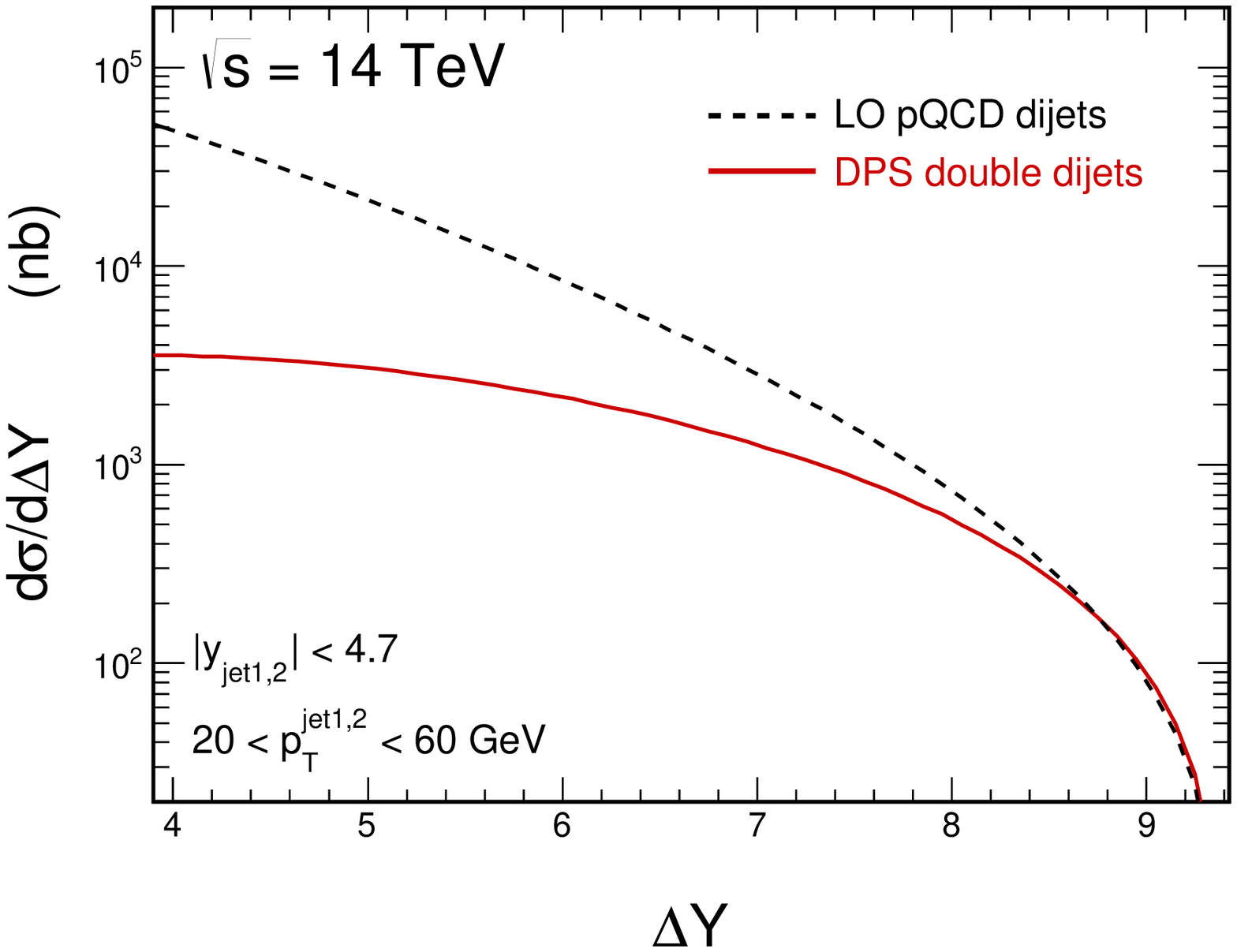} \\
\includegraphics[width=4cm]{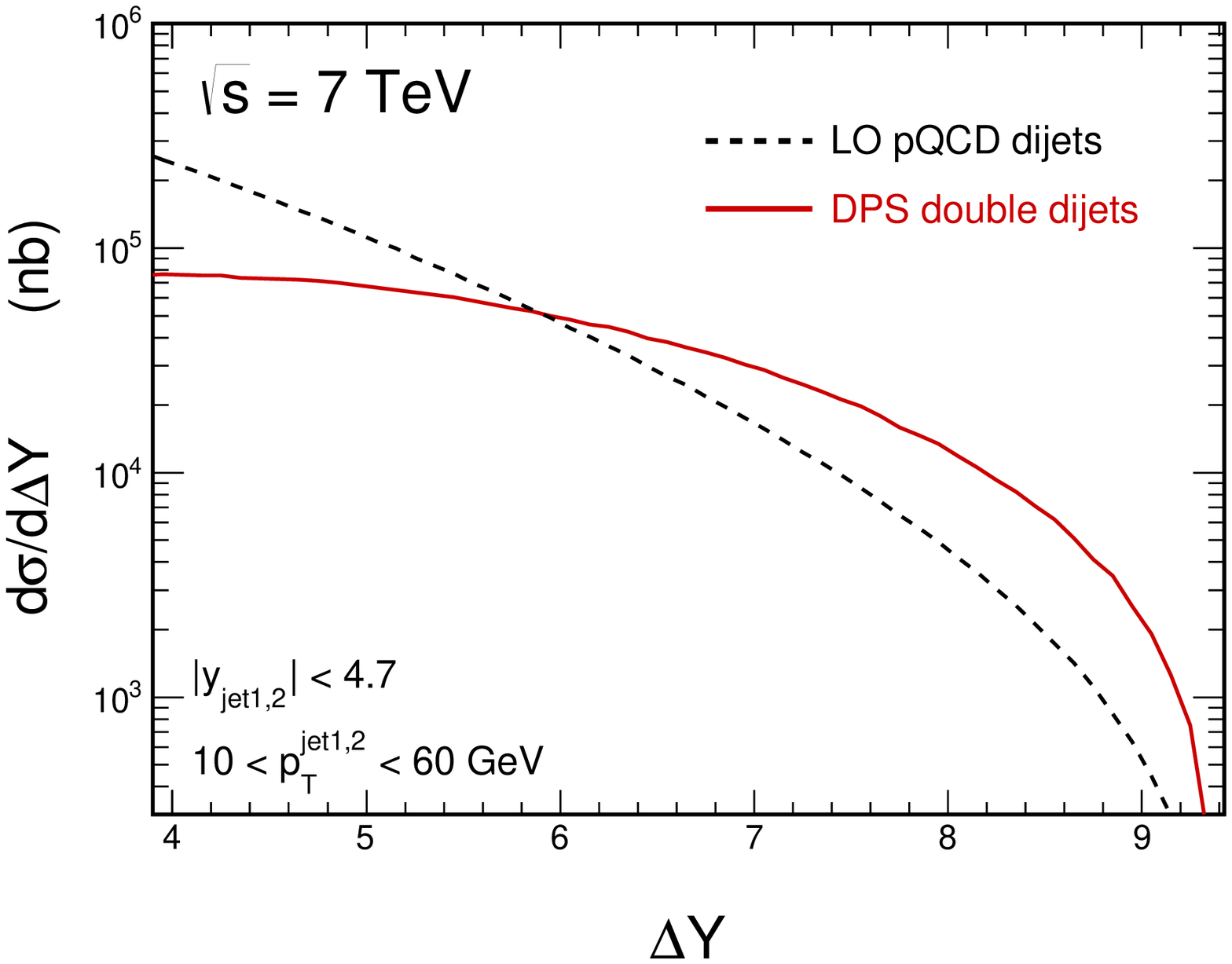}
\includegraphics[width=4cm]{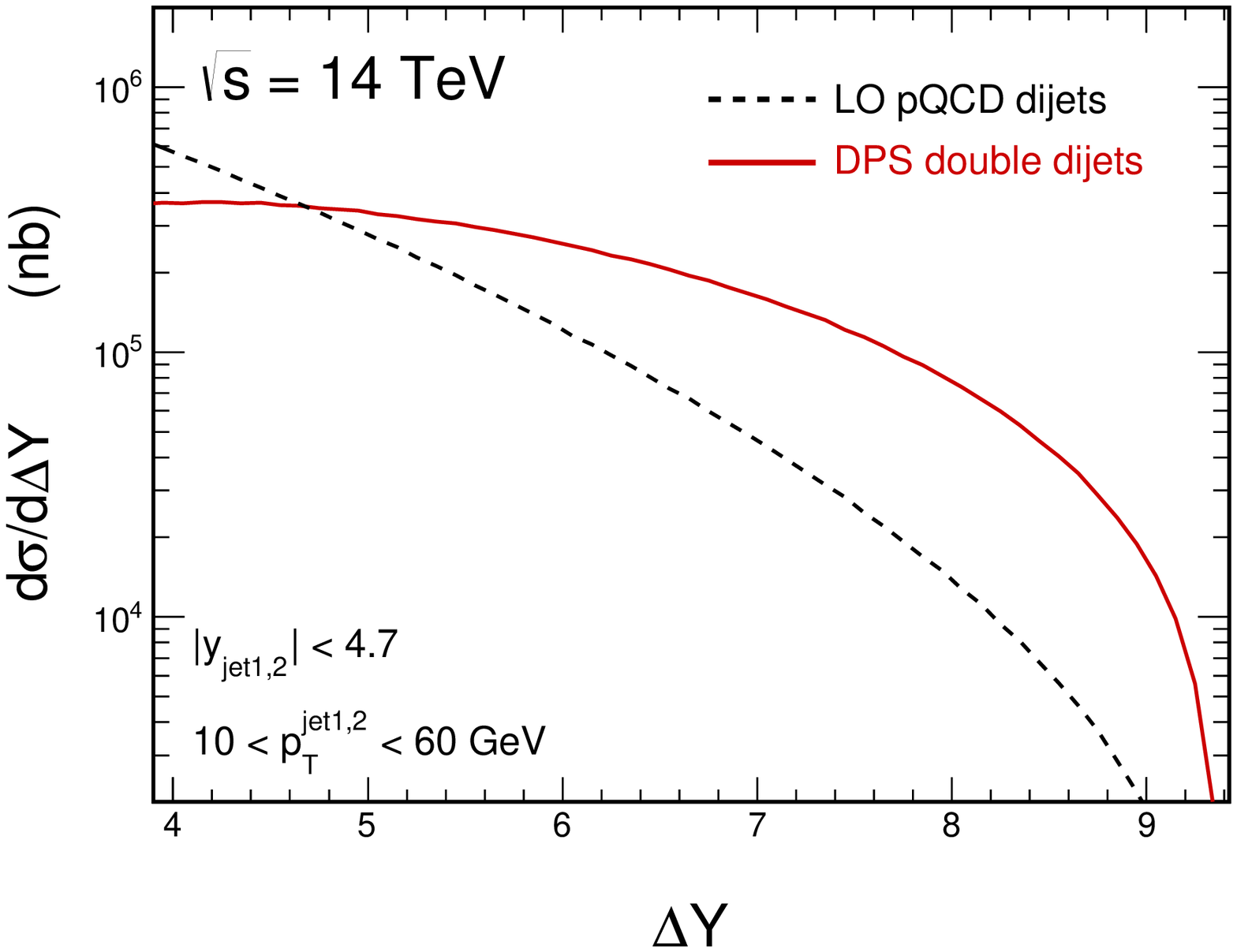}
   \caption{
\small The same as in the previous figure but now for somewhat smaller lower
cut on minijet transverse momentum.
}
 \label{fig:Deltay-2}
\end{figure}

Let us now show results for even smaller transverse momenta
(see Fig.~\ref{fig:Deltay-2}). 
A measurement of such minijets may be, however, difficult. 
Now the DPS contribution may even exceed the standard SPS 
dijet contribution. 
How to measure such minijets is an open issue. In principle
one could measure for instance correlations of 
semihard ($p_t \sim$ 10 GeV) neutral pions with the help of 
so-called zero-degree calorimeters (ZDC). This will be discussed elsewhere.

\begin{figure}[!h]
\includegraphics[width=5.5cm]{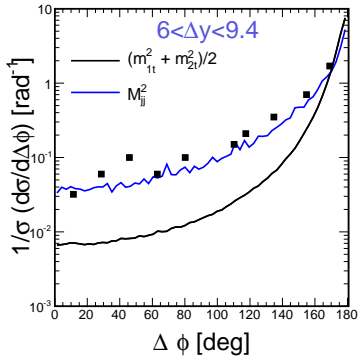}
  \caption{
\small Dijet azimuthal angle distribution for large rapidity separation.
We show result of $k_t$-factorization approach \cite{CMS2014}
with Kimber-Martin-Ryskin unintegrated distributions for different 
factorization scales as well as preliminary experimental data of the 
CMS collaboration \cite{CMS_MN1}.
}
 \label{fig:dsig_dphi}
\end{figure}

Finally let us discuss azimuthal correlations between jets widely
separated in rapidity.
In Fig.\ref{fig:dsig_dphi} we show results obtained in 
the $k_t$-factorization approach with the Kimber-Martin-Ryskin
UGDF \cite{KMR} for different factorization scales described in the figure.
The distributions are normalized to the integrated cross section
for the corresponding experimental acceptance range.
In fact the cross section obtained with $\mu_f^2 = M_{jj}^2$ is much
smaller than that obtained with $\mu_f^2 = m_t^2$ 
($m_t$ is jet transverse mass) \cite{CMS2014}.
The distribution corresponding to DPS is a straight line (the two different
hard processes are not correlated) and 
considerably modify the distribution for $\mu_f^2 = M_{jj}^2$
and only slightly modify the distribution corresponding to
$\mu_f^2 = m_t^2$. 
All the details of the calculations will be presented in our future
paper \cite{CMS2014}.

\section{Conclusions and outlook}

In this presentation we have discussed how the double-parton scattering
effects may contribute to large-rapidity-distance dijet correlations.
The calculation was performed in leading-order
to understand and explore the general situation. This means that 
each step of DPS was calculated in pQCD leading order. We have shown 
that already leading-order calculation provides quite adequate
description of inclusive jet production when confronted with
the results obtained by the ATLAS collaboration.

We have concentrated on distributions in rapidity distance between
the most-distant jets in rapidity. The results of the dijet SPS
mechanism have been compared to those for the DPS mechanism. We have performed
calculations relevant for a planned CMS analysis. The contribution of 
the DPS mechanism increases with increasing distance in rapidity between
jets.

We have also shown some recent predictions of the Mueller-Navelet jets
in the LL and NLL BFKL framework.
For the CMS configuration the DPS contribution is smaller than 
the dijet collinear SPS contribution even at the high rapidity distances
and only slightly smaller than that for the NLL BFKL calculation.
The DPS final state topology is clearly different than that for the
collinear dijet SPS which may help to disentangle the two mechanisms.
More refined analysis should take into account two- three- and four-jet 
final states.

We have shown that the relative effect of DPS could be increased
by lowering the transverse momenta of jets but such measurements
can be difficult if not impossible (in addition there is a similar
tendency in the BFKL approach).
Alternatively one could study correlations of semihard neutral pions 
widely separated in rapidity.
LHCf collaboration at ATLAS could consider such measurements. 

The DPS effects are interesting not only in the context how they 
contribute to distribution in rapidity distance between jets but also 
per se.
One could make use of correlations in jet transverse momenta,
jet transverse-momentum imbalance (see \cite{MS2014}) and 
azimuthal correlations to enhance the contribution of DPS. 

\bibliographystyle{aipprocl} 


\end{document}